\documentclass{jfm}
\usepackage{graphicx}
\usepackage{epstopdf, epsfig}

\usepackage{xcolor}
\usepackage{comment}

\newcommand\Web{\mbox{\textit{We}}} 
\newcommand\Deb{\mbox{\textit{De}}} 

\shorttitle{What determines the drop size in sprays of polymer solutions?}
\shortauthor{A. Gaillard, R. Sijs and D. Bonn}

\title{What determines the drop size in sprays of polymer solutions?}

\author{A. Gaillard\aff{1}
  \corresp{\email{antoine0gaillard@gmail.com}},
  R.Sijs\aff{1}
 \and D.Bonn\aff{1}}

\affiliation{\aff{1}Van der Waals-Zeeman Institute, University of Amsterdam, Science Park 904, Amsterdam, Netherlands}

\begin{document}

\maketitle

\begin{abstract}
The effect of viscoelasticity on sprays produced from agricultural flat fan nozzles is investigated experimentally using dilute aqueous solutions of polyethylene oxide (PEO). Measurements of the droplet size distribution using laser diffraction reveal that polymer addition to water results in the formation of overall bigger droplets with a broader size distribution. The median droplet size $D_{50}$ is found to increase linearly with the extensional relaxation time of the liquid. The non-dimensional median droplet sizes of different polymer solutions, sprayed at different operating pressures from nozzles of different sizes, rescale on a single master curve when plotted against an empirical function of the Weber and Deborah numbers. Using high-speed photography of the spraying process, we show that the increase in droplet size with viscoelasticity can be partly attributed to an increase of the wavelength of the flapping motion responsible for the sheet breakup. We also show that droplet size distributions, rescaled by the average drop size, are well described by a compound gamma distribution with  parameters $n$ and $m$ encoding for the ligament corrugation and the width of the ligament size distribution, respectively. These parameters are found to saturate to values $n=4$ and $m=4$ at high polymer concentrations.\\
\end{abstract}


\section{Introduction}


Spraying is a widely used technique that involves the fragmentation of a liquid sheet or jet into droplets. This process usually involves multiple hydrodynamic instabilities  \citep{villermaux2007fragmentation,dombrowski1954photographic}. Sprays are important to a wide range of application areas such as agriculture (crop spraying), industry (inkjet printing, spray painting) and medicine (drug delivery) \citep{lefebvre2017atomization,bayvel1993liquid}. In many applications, understanding how the size distribution of spray droplets depends on external control parameters is of paramount importance. This is particularly true for agricultural pesticide spraying \citep{hislop1987can} where droplets typically smaller than 100 micrometers may be blown away by the wind (spray drift), causing environmental pollution \citep{reichenberger2007mitigation, stainier2006droplet, matthews2008pesticide}. Addition of small amounts of polymer molecules to the sprayed liquid has been shown to lead to the formation of larger droplets, hence reducing the volume of spray drift produced by agricultural nozzles \citep{stelter2002influence,williams2008influence}, while also improving droplet deposition by inhibiting the bouncing effect observed for water droplets \citep{bartolo2007dynamics,smith2010effect,xu2021quantifying}. Polymers hence offer advantages over surfactant additives, which have been shown to only improve the deposition process \citep{hoffman2021controlling} and not the droplet size distribution \citep{sijs2020effect}. Understanding the effect of viscoelasticity in liquid atomization is also relevant for jet fuel and fire-fighting applications \citep{chao1984antimisting,hoyt1974structure}, as well as for understanding virus propagation through sneezing \citep{scharfman2016visualization}. However, viscoelastic effects in sprays are not yet fully understood \citep{villermaux2020fragmentation}. 

\begin{figure}
  \centerline{\includegraphics[scale=1.2]{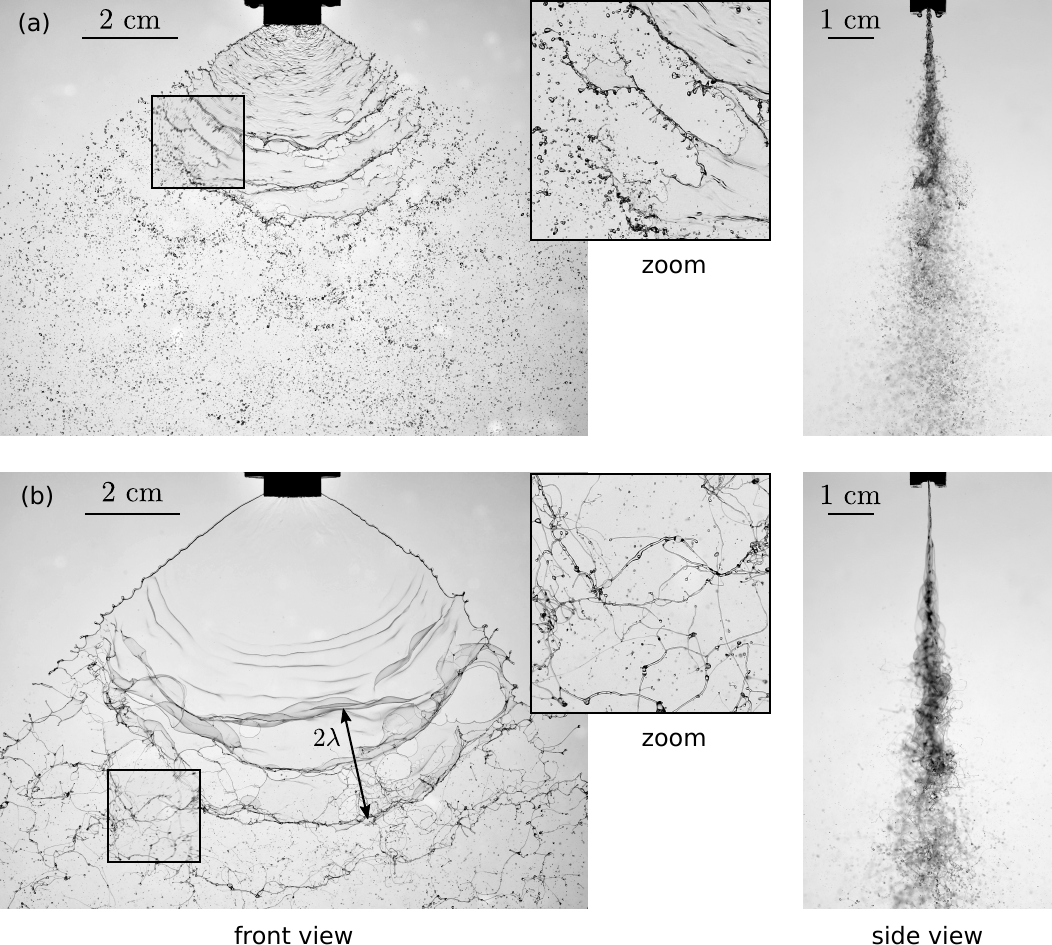}}
  \caption{Front view (left) and side view (right) images of a spray of water (\textit{a}) and of 25\,ppm PEO-4M solution (\textit{b}), obtained using high-speed photography. The sprays are generated by a Teejet 110-04 flat fan nozzle at the same operating pressure of 2 bar. Inset images labelled `zoom' correspond to the highlighted sections of the front-view images. The arrow in the left figure of panel (b) shows a typical estimation of the Squire wavelength discussed in \S\ref{sec:Squire wavelength}.}
\label{fig:spray_images}
\end{figure}

In this paper, we focus on the fragmentation of liquid sheets generated by flat fan nozzles. For dilute solutions, the mechanism leading to the formation of spray droplets is qualitatively similar to that of Newtonian liquids \citep{stelter2002influence,hartranft2003sheet}. This is illustrated in figure \ref{fig:spray_images} which compares front-view and side-view images of spray sheets of water (a) and of a dilute solution of high-molecular-weight polyethylene oxide (PEO). The fragmentation of the spray sheet into droplets is initiated by the growth of out-of-plane oscillations along the sheet caused by friction with the surrounding air via the Squire instability \citep{squire1953investigation}, as can be observed in the side-view images of figure \ref{fig:spray_images}. Thickness modulations in the sheet caused by this flapping motion result in the rupture of the sheet into fragments of size $\lambda$ corresponding to the Squire wavelength \citep{dombrowski1963aerodynamic}, as can be observed in the front-view images of figure \ref{fig:spray_images}. These sheet fragments retract into ligaments which ultimately breakup into droplets due to surface tension via the Rayleigh-Plateau instability \citep{rayleigh1878instability}. 

\citet{kooij2018determines} derived a model for low-viscosity Newtonian liquids where the median diameter $D_{50}$ of spray droplets is set by the Squire wavelength and by the thickness of the sheet at the point of breakup. This model leads to the following universal expression
\begin{equation}
D_{50}/b = c_1 \alpha^{-1/6} \Web^{-1/3}
\label{eq:Kooij}
\end{equation}
where $\alpha$ is the air-to-liquid-density ratio, $c_1$ is a numerical constant and the Weber number is $\Web = \rho U^2 b / \sigma$ with $b$ is a characteristic size of the nozzle opening, $U$ the liquid velocity, $\rho$ the liquid density and $\sigma$ the liquid surface tension. This expression, which doesn't account for viscous effects, was found to be in agreement with experimental measurements for liquids of viscosities up to 30 times larger than that of water. 

Many studies have shown that increasing the extensional viscosity of the sprayed liquid by adding polymers to a Newtonian solvent leads to the formation of larger droplets, for both flat fan nozzles and pressure swirl atomizers producing flat and conical sheets, respectively \citep{dexter1996measurement,zhu1997effects,mun1999atomisation,williams2008influence,park2008effects}. The first attempt to propose a correlation between the characteristic droplet size and pertinent non-dimensional numbers is found in the work of \citet{stelter2002influence}, who performed measurements of the droplet size distribution for different nozzles, spraying dilute aqueous solutions of both flexible and rigid polymers. Viscoelasticity is quantified by an extensional relaxation time $\tau$ measured using a Capillary Breakup Extensional Rheometer (CaBER). Their empirical expression predicts that the (Sauter) mean diameter, expressed in terms of Weber and Reynolds numbers, scales as $\tau^{0.425}$ for viscoelastic liquids and scales as $\eta^{0.213}$ for Newtonian liquids of dynamic shear viscosity $\eta$. However, these predictions are inconsistent with the absence of viscous effects measured by \citet{kooij2018determines} for low-viscosity Newtonian liquids. A more recent attempt is found in the work of \citet{brenn2017formation}, who used a linear analysis of the instability of viscoelastic liquid sheets \citep{liu1998linear} and jets \citep{liu2006linear} to derive an expression in which the increase in droplet size after polymer addition is mainly attributed to the increase of shear viscosity of the liquid, while an increase in polymer relaxation time for a given shear viscosity leads to a decrease in droplet size. These predictions are inconsistent with the increase in droplet size observed by \citet{mun1999atomisation} for polymer solutions exhibiting the same shear viscosity. In fact, these predictions are based on a linearized viscoelastic model which do not account for the strain hardening properties associated with large deformations of polymer molecules in spray flows. 

Hence, to the best of our knowledge, no satisfactory expression has been proposed for the characteristic droplet size in sprays of viscoelastic liquids produced by flat fan nozzles. For air-assisted jet atomization nozzles on the other hand, \citet{keshavarz2015studying} showed a logarithmic increase of the mean droplet size with the relaxation time of the polymer solution, consistent with  earlier measurements by \citet{christanti2006quantifying}. \citet{keshavarz2015studying} also proposed a phenomenological model that accounts for their observations. 

For flat fan nozzles, the physical origin of the increase in droplet size with increasing viscoelasticity is also an open question. \citet{stelter2002influence} proposed that it is a consequence of the retardation of ligament breakup caused by strain hardening. The authors also proposed that the ligament size is not affected since the sheet thickness at breakup was estimated to be unchanged after polymer addition. Later, \citet{hartranft2003sheet} proposed that spray droplets are larger because of an increase in ligament size caused by (i) an increase in the Squire wavelength and (ii) a decrease in sheet length (while \citet{stelter2002influence} reported an increase in the sheet length in their experiments). However, these studies did not provide quantitative comparisons of the drop size against relevant physical quantities to validate their hypothesis. 

In this study, we propose a universal empirical expression of the median diameter $D_{50}$ of spray droplets of polymer solutions generated from flat fan nozzles in terms of the non-dimensional Weber and Deborah numbers. This expression, which reduces to equation (\ref{eq:Kooij}) for vanishing viscoelastic effects, is shown to successfully capture measurements performed for dilute aqueous solutions of PEO sprayed at different operating pressure from nozzles of different sizes. The physical origin of the increasing drop size with increasing liquid elasticity is also investigated via measurements of different quantities involved in the sheet fragmentation mechanism, revealing a significant role of the increasing Squire wavelength. Finally, the effect of viscoelasticity on the droplet size distributions is discussed.

Section \ref{sec:Materials and methods} describes the experimental setup and the measurement techniques as well as the rheology of the test liquids. It also includes a discussion of the mechanical degradation of the polymer solutions. The experimental results are presented in section \ref{sec:Results and discussions} before conclusions are drawn in section \ref{sec:Conclusions}.

\section{Materials and methods}
\label{sec:Materials and methods}

\subsection{Nozzles}
\label{sec:Nozzles}

Sprays are generated using two flat fan nozzles provided by TeeJet\textsuperscript{\textregistered} Technologies, producing flat liquid sheets. The dimensions of the oval opening of each nozzle, measured from high-resolution images, are reported in table \ref{tab:nozzles}, where $A$ is the nozzle area and $b$ and $a$ are the nozzle thickness and width corresponding to the minimum and maximum lengths of the cross section area, respectively. The spray angle $\theta_w$ measured for water from front pictures of the sheet is observed to increase in a range reported in table \ref{tab:nozzles} when increasing the operating pressure from 1 to 8 bar.

\begin{table}
\setlength{\tabcolsep}{3pt}
  \begin{center}
\def~{\hphantom{0}}
  \begin{tabular}{ccccccc}
  
      Type & Nozzle & $A$ (mm$^2$) & $b$ (mm) & $a$ (mm)   &  $a/b$ & $\theta_w$ \\[8pt]
     
       Flat fan & XR Teejet 11004 VS  & 1.15  & 0.601 & 2.25 & 3.74 & 105 - 120 \\
       Flat fan & XR Teejet 11002 VK  & 0.545 & 0.458 & 1.52 & 3.31 & ~93 - 114  \\

  \end{tabular}
  \caption{Dimensions of the oval nozzle opening of the two flat fan nozzles: area $A$, thickness $b$, width $a$, aspect ratio $a/b$. The range of spray angles $\theta_w$ measured for water corresponds to operating pressures from 1 to 8 bar.} 
  \label{tab:nozzles}
  \end{center}
\end{table}

\subsection{Test liquids and injection system}
\label{sec:Test liquids and injection system}

Spray experiments are performed using aqueous solutions of polyethylene oxide with molecular weight $M_w = 4000$~kg/mol provided by Alrokon\textsuperscript{\textregistered}, henceforth referred to as PEO-4M. Solutions are obtained from dilution of 10000\,ppm (weight parts per million) stock solutions mixed for $16$ hours by a mechanical stirrer to ensure homogeneity. To minimize aggregation, a mass of isopropanol equal to the mass of polymer is added to the polymer powder before it is added to water. Rheological measurements reveal that stock solutions prepared via this protocol are highly reproducible. 

The polymer solutions fed to the nozzle are generated by a pressure powered polymer injection system from Dosatron\textsuperscript{\textregistered} (ref. D25WL2 IE PO + Mixer). Tap water is pumped into a water inlet, generating suction of a concentrated polymer solution through a polymer inlet, which is injected into the water flow at a controlled volume ratio. Polymer injection is followed by passage through a mixing chamber delivering a homogeneous diluted polymer solution. The concentration of these diluted solutions is varied between 10 and 100~ppm. The mixer outlet is connected to the nozzle inlet through a rigid pipe equipped with a pressure and flow rate meter provided by Krohne\textsuperscript{\textregistered} (ref. OPTIBAR PM 3050 and H250 RR M40) and time measurements of the flow rate and operating pressure are recorded on a computer. The Dosatron injection system causes periodic pressure variations of the order of 0.2~bar on a time scale of about 20 seconds. Hence, the experimental data reported in this work correspond to averages calculated over 2 minutes of experiments performed at a constant average pressure $P$ and volumetric flow rate $Q$. 

The flow rate varies with pressure according to $Q = C_d A \sqrt{2P/\rho}$, where $\rho$ is the liquid density and $C_d$ a discharge coefficient, which is found to not vary significantly with polymer concentration and is close to $1$, as also reported by \citet{kooij2018determines}. The operating pressure is varied from 1 to 8 bar, yielding liquid velocities $U_0 = Q/A$ in the nozzle ranging between 10 and 40~m/s for both nozzles.

\subsection{Droplet size measurements}
\label{sec:Droplet size measurements}

The size distribution of the spray droplets is measured by a laser diffraction instrument from Malvern Spraytech\textsuperscript{\textregistered}. This method has been shown to give results comparable to other measurements techniques \citep{sijs2021drop}. The laser beam is placed vertically below the nozzle and passes through the spray in a direction perpendicular to its plane. The diffraction angle being inversely proportional to the droplet size, the light diffraction pattern allows, assuming a spherical shape of the droplets, to obtain the droplet size distribution \citep{swithenbank1976laser, dayal2004evaluation}. The laser beam is placed at a fixed distance 50~cm below the nozzle where high-speed photography reveals that droplets are reasonably spherical and no longer connected by filaments, see \citet{christanti2006quantifying} for calculations of the effect of filaments on drop size measurements.

\subsection{Shear rheology}
\label{sec:Shear rheology}

In \S\ref{sec:Shear rheology} and \S\ref{sec:Extensional rheology}, we present shear and extensional rheology measurements performed on PEO solutions prepared by dilutions from a 10000~ppm stock solution. These measurements include concentrations above the maximum concentration 100~ppm used in spray experiments for a more general characterization of aqueous PEO solutions. The rheology of solutions used in spray experiments is discussed in \S\ref{sec:Mechanical degradation}. Rheological measurements are performed at the temperature of spray experiments ($18.5^{\circ}$C).

\begin{figure}
  \centerline{\includegraphics[scale=1]{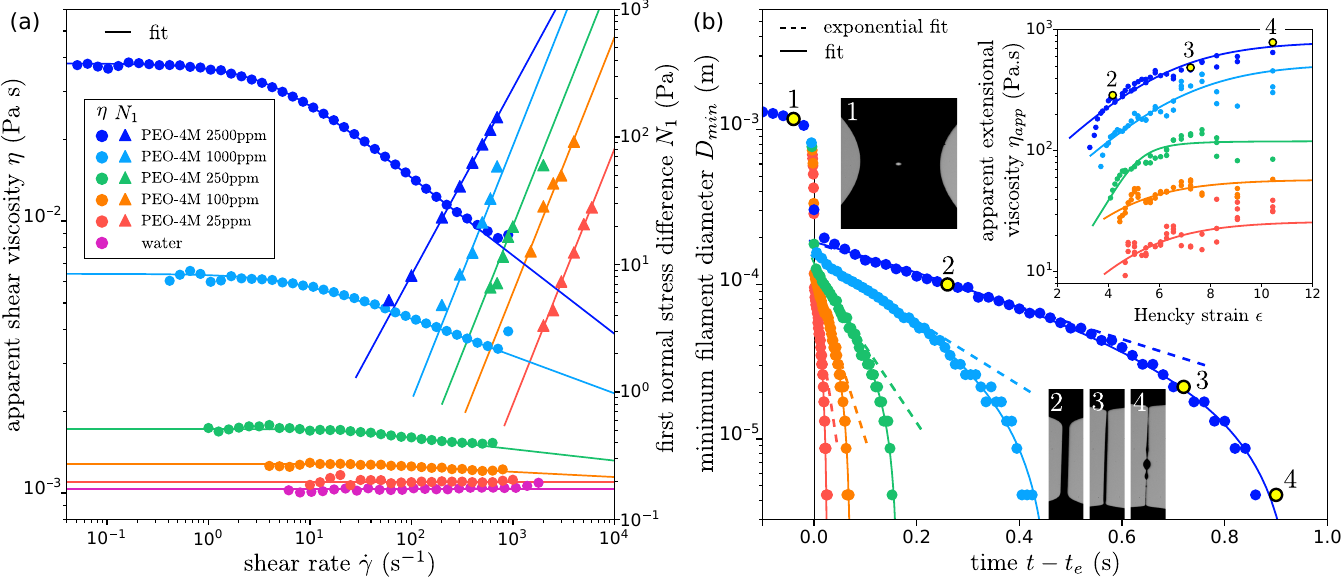}}
  \caption{(a) Apparent shear viscosity $\eta$ ($\bullet$) and first normal stress difference $N_1$ ($\blacktriangle$) as a function of shear rate $\dot{\gamma}$ for water and PEO-4M polymer solutions of various concentrations. Data are fitted respectively by equations (\ref{eq:carreau}) and (\ref{eq:N1}). (b) Minimum filament diameter $D_{min}$ as a function of the time $t-t_e$ elapsed since the beginning of the elastic regime for the same solutions, with the same colour coding as panel (a). Data are fitted by the empirical law proposed by \citet{anna2001elasto}: $D_{min} = a e^{-b t} - c t + d$, where $a,b,c$ and $d$ are fitting parameters. The inset figure shows the apparent extensional viscosity $\eta_{app} = - \sigma / (\mathrm{d} D_{min}/\mathrm{d}t)$ as a function of the Hencky strain $\epsilon = -2\ln(D_{min}/D_1)$, where we choose $D_1 = 1$~mm. Inset snapshots labelled 1 to 4 show a filament of a 2500 ppm solution at different stages of thinning indicated by the same labels on the corresponding $D_{min}$ and $\eta_{app}$ curves.}
\label{fig:rheology}
\end{figure}

Shear rheology measurements are performed using a MRC-302 rheometer from Anton Paar equipped with a cone plate geometry (diameter 50~mm, angle $1^{\circ}$ and truncation gap $53$~$\mu$m). The apparent shear viscosity $\eta$ and first normal stress difference $N_1$ are plotted in figure \ref{fig:rheology}(a) against the shear rate $\dot{\gamma}$ for water and PEO-4M solutions. To measure $N_1$, we follow a step-by-step protocol similar to \citet{casanellas2016stabilizing} in order to circumvent the instrumental drift of the normal force. This protocol consists in applying steps of constant shear rate followed by steps of zero shear and subtracting the two raw $N_1$ plateau values. The contribution of inertia to the normal force is corrected for by the rheometer \citep{macosko1994rheology}. 

We fit the apparent shear viscosity in figure \ref{fig:rheology}(a) using the Carreau-Yasuda formula:
\begin{equation}
\eta(\dot{\gamma}) = \eta_0 ( 1 + ( \dot{\gamma}/\dot{\gamma}_c)^{a_1})^{(n-1)/a_1},
\label{eq:carreau}
\end{equation}
where $\eta_0$ is the zero shear viscosity, $\dot{\gamma}_c$ is the shear rare beyond which shear thinning is observed, $n$ is the degree of shear thinning and $a_1$ encodes the sharpness of the transition towards the shear thinning regime. 
The first normal stress difference $N_1$ in figure \ref{fig:rheology}(a) can be described by a power law:
\begin{equation}
N_1 = \Psi_1 \dot{\gamma}^{\alpha_1},
\label{eq:N1}
\end{equation}
where $\Psi_1$ is the first normal stress coefficient. All fitting parameters are reported in table \ref{tab:rheology}. The table also includes the surface tension $\sigma$ of the solutions, which is measured by a pendant drop method. The liquid density $\rho$ is 997~kg/m$^3$ for all solutions.

The polymer viscosity is defined as $\eta_p = \eta_0 - \eta_s$ where $\eta_s = 1.036$~mPa~s is the viscosity of the water solvent. We find that $\eta_p$ increases linearly at low polymer concentrations (dilute regime), i.e. $\eta_p = \eta_s [\eta] c$ with $c$ the polymer mass concentration. We find an intrinsic viscosity $[\eta] = 2.61$~m$^3$/kg. Following the expression of \citet{graessley1980polymer}, the critical overlap concentration is $c^* = 0.77/[\eta] = 0.295$~kg/m$^3$ (equivalently 296~ppm). Hence, PEO solutions used in spray experiments with concentrations up to 100~ppm are dilute solutions with a quasi constant shear viscosity $\eta_0$ ($n \approx 1$) and quadratic first normal stress difference ($\alpha_1 = 2$), see table \ref{tab:rheology}.

\begin{table}
\setlength{\tabcolsep}{2.5pt}
  \begin{center}
\def~{\hphantom{0}}
  \begin{tabular}{ccccccccccccc}
  
      [PEO] & $c/c^*$ & $\sigma$ & $\eta_0$   &  $\eta_p$  & $n$ & $1/\dot{\gamma}_c$  & $\alpha_1$ & $\Psi_1$  & $\tau$ & $\eta_E$ & $L^2$  \\
      
       (ppm) & & (mN/m) & (mPa~s)  &  (mPa~s) &  & (s) &  & (Pa~s$^{\alpha_1}$) & (ms)   & (Pa~s) & \\ [8pt]
      
     
       10~~~  & 0.034 & 67.1 & 1.037 & 0.027  & 1.00  & --      & --   & --                   & 1.16  & 5.3~   & 9.8$\times 10^4$\\
       25~~~  & 0.084 & 64.1 & 1.098 & 0.062  & 1.00  & --      & 2    &   1$\times 10^{-6}$  & 5.95  & 18~~   & 1.5$\times 10^5$\\
       50~~~  & 0.17~ & 63.0 & 1.17~ & 0.134  & 0.98  & 0.0258  & 2    & 2.5$\times 10^{-6}$  & 6.54  & 25~~   & 9.3$\times 10^4$\\
       100~~  & 0.34~ & 62.8 & 1.28~ & 0.244  & 0.98  & 0.0335  & 2    & 6.0$\times 10^{-6}$  & 13.8  & 45~~   & 9.2$\times 10^4$\\
       150~~  & 0.51~ & 62.5 & 1.44~ & 0.404  & 0.97  & 0.0445  & 2    & 8.0$\times 10^{-6}$  & 18.5  & 70~~   & 8.7$\times 10^4$\\
       250~~  & 0.84~ & 62.5 & 1.72~ & 0.684  & 0.96  & 0.0575  & 2    & 2.0$\times 10^{-5}$  & 28.5  & 130~   & 9.5$\times 10^4$\\
       500~~  & 1.7~~ & 62.5 & 2.95~ & 1.91~  & 0.93  & 0.120~  & 2    & 5.5$\times 10^{-5}$  & 49.0  & 190~   & 5.0$\times 10^4$\\
       1000~  & 3.4~~ & 62.5 & 6.3~~ & 5.26~  & 0.86  & 0.152~  & 2    & 8.5$\times 10^{-5}$  & 70.0  & 270~   & 2.6$\times 10^4$\\
       2500~  & 8.4~~ & 62.1 & 38~~~ & 37.0~  & 0.71  & 0.282~  & 1.47 & 9.6$\times 10^{-3}$  & 136  & 530~   & 7.2$\times 10^3$\\
       10000  & 34~~  & 61.5 & 15000 & 15000  & 0.42  & 23.8~~  & 1.05 & 2.2$\times 10^{0}$~  & 508  & 1400  & 4.7$\times 10^1$\\

  \end{tabular}
  \caption{Rheological parameters and surface tension $\sigma$ of PEO solutions used for rheological characterization. Shear parameters: $\eta_0$, $n$, $\dot{\gamma}_c$, $\alpha_1$ and $\Psi_1$ are such that the shear viscosity $\eta (\dot{\gamma})$ and the first normal stress difference $N_1(\dot{\gamma})$ are captured by equations (\ref{eq:carreau}) and (\ref{eq:N1}) ($N_1$ was not measurable at 10~ppm). Values of $c/c^*$ are indicated, where $c$ is the polymer mass fraction and $c^* = 0.295$~kg/m$^3$ (equivalently 296~ppm) is the critical overlap concentration. $\eta_p = \eta_0 - \eta_s$, where $\eta_s = 1.036$~mPa~s is the water solvent viscosity. We use $\eta_p = \eta_s [\eta] c$ to determine $\eta_p$ at 10~ppm since the difference between $\eta_0$ and $\eta_s$ is too small to be measured directly, where $[\eta] = 2.61$~m$^3$/kg is the intrinsic viscosity. Extensional parameters (CaBER): $\tau$ is the extensional relaxation time and $\eta_E$ is the terminal extensional viscosity. $L^2 = (\eta_E - 3 \eta_s)/(2 \eta_p)$ is the effective value of the finite extensibility parameter.} 
  \label{tab:rheology}
  \end{center}
\end{table}

\subsection{Extensional rheology}
\label{sec:Extensional rheology}

Extensional rheology measurements are performed using a non-commercial Capillary Breakup Extensional Rheometer (CaBER) based on the filament thinning technique documented by \citet{anna2001elasto}. A drop of polymer solution is placed between two horizontal plates which are separated until the liquid bridge connecting the two end drops becomes unstable, transiently forming a filament in which polymer molecules are stretched by an extensional flow. Following \citet{campo2010slow}, the plates are separated at a low velocity ($200$~$\mu$m/s) to minimize oscillations of the two end drops inherent to the classical step-strain plate separation protocols \citep{rodd2004capillary}. The minimum filament diameter $D_{min}$ is calculated from images recorded by a high-speed camera equipped with a high-magnification objective.

Time evolutions of the minimum filament diameter are shown in figure \ref{fig:rheology}(b) for PEO-4M solutions of various concentrations, showing an exponential decay in the elastic regime. This is consistent with the multimode Oldroyd-B model, which for the elastic regime (starting at $t=t_e$) predicts \citep{bazilevskii1997failure,entov1997effect,clasen2006dilute}:
\begin{equation}
D_{min}(t) = \left( \frac{G D_0^4}{4 \sigma}\right)^{1/3} \exp{\left(- \frac{t-t_e}{3 \tau} \right)},
\label{eq:D_filament}
\end{equation}
where $G$ is the elastic modulus, $D_0$ the minimum bridge diameter at the onset of instability and $\tau$ the longest relaxation time of the spectrum. Values of $\tau$ are obtained from fitting the experimental exponential regime. In this regime, the extension rate $\dot{\epsilon} \equiv - 2(\mathrm{d}D_{min}/\mathrm{d}t) / D_{min}$ gives $\tau \dot{\epsilon} = 2/3$ which is above the coil-stretch transition value $1/2$ \citep{de1974coil}. Hence, polymer chains progressively unravel, ultimately reaching full extension, at which point the apparent extensional viscosity defined as 
\begin{equation}
\eta_{app} = - \sigma / ( \mathrm{d}D_{min}/\mathrm{d}t), 
\label{eq:etaapp}
\end{equation}
saturates to a value $\eta_E$ called the terminal extensional viscosity. The FENE-P model predicts $\eta_E = 3 \eta_s + 2 \eta_p L^2$ \citep{mckinley2005visco} where $L^2$ is the finite extensibility parameter and $L$ is the ratio of the polymer size at full extension to its size in the coiled state at equilibrium. Values of $\eta_E$ are obtained by plotting $\eta_{app}$ as a function of the Hencky strain $\epsilon = -2 \ln{(D_{min}/D_1)}$, where $D_1$ is the minimum bridge diameter slightly before onset on the elastic regime when polymer molecule start unravelling, see inset of figure \ref{fig:rheology}(b). Values of $\tau$, $\eta_E$ and $L^2 = (\eta_E - 3 \eta_s)/(2 \eta_p)$ are reported in table \ref{tab:rheology}. We find $\tau \propto c^{0.76}$, consistent with \citet{zell2010there} and $\eta_E \propto \tau$, consistent with \citet{stelter2002investigation}.

\subsection{Mechanical degradation}
\label{sec:Mechanical degradation}

\begin{figure}
  \centerline{\includegraphics[scale=1]{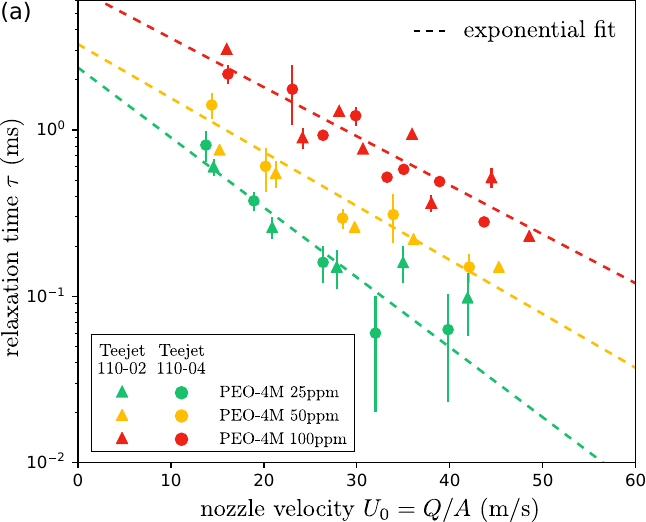}}
  \caption{Extensional relaxation time $\tau$ of sprayed liquids after mechanical degradation, measured using the CaBER technique, as a function of the liquid velocity. Sprays are generated using Teejet 110-04 ($\blacktriangle$) and 110-02 ($\bullet$) nozzles with opening area $A$ at different flow rates $Q$. Liquid velocity $U_0 = Q/A$. Dotted lines are exponential fits.}
\label{fig:degradation}
\end{figure}

The rheology of the polymer solutions used in spay experiments is investigated by collecting samples of the of the sprayed liquid. In figure \ref{fig:degradation}, we plot the extensional relaxation time $\tau$ measured on samples of different polymer concentrations, sprayed at different flow rates from both nozzles, against the liquid velocity $U_0$ in the nozzle. A significant reduction of the relaxation time is observed when increasing the flow rate, meaning that polymer solutions undergo severe mechanical degradation, as also reported by \citet{stelter2002influence} and \citet{hartranft2003sheet}. Since the spray formation is governed by the properties of the liquid in the spray sheet, it would be incorrect to correlate spray properties to the rheology of the initial fresh solution before degradation. Figure \ref{fig:degradation} shows that the decrease in $\tau(U_0)$ is well fitted by an exponential decay for every polymer concentration. The parameters of the fits are hence used to calculate the relaxation time in the spray experiments reported in \S\ref{sec:Results and discussions}. 

We note that values of $\tau$ in figure \ref{fig:degradation} are the same for both nozzles at a given liquid velocity and polymer concentration. This is surprising since a reduction of the nozzle area is expected to lead to more degradation, owing to the increase in strain rate associated to the contraction flow at the nozzle inlet, as reported by \citet{stelter2002influence}. We therefore suspect that degradation might not only occur at the nozzle opening in our experiments.

\section{Results and discussions}
\label{sec:Results and discussions}

We start by a qualitative description of the sheet fragmentation process. We find that, similarly to Newtonian fluids, the fragmentation of the viscoelastic spray sheet is initiated by the growth of Squire waves along the sheet, leading to the formation of sheet fragments retracting into ligaments which ultimately breakup into droplets. This is shown in the high-speed images of figure \ref{fig:spray_images}, which compares a spray of water (a) and of 25~ppm PEO-4M solution for the same nozzle and operating pressure. However, significant qualitative differences are observed in these images. 

First, the liquid sheet close to the nozzle is much smoother in the viscoelastic case, reminiscent of the reduction of small-scale surface disturbances in jets of polymer solutions reported by \citet{hoyt1974structure}. The rim of the sheet is also stabilized, as also observed by \citet{thompson2007atomization} for flat fan sprays of wormlike micelle solutions. We also observe that polymers delay the breakup of the sheet to a larger distance away from the nozzle, consistent with observations by \citet{stelter2002influence}, suggesting a stabilizing effect of viscoelasticity. This is reminiscent of the work of \citet{karim2018effect}, who found that polymer addition may prevent the breakup of a liquid curtain subjected to a localized perturbation created by an air jet. The authors concluded that the growth rate of disturbances that may lead to a perforation of the sheet is delayed by elastic stresses. Following sheet fragmentation, we observe that the breakup of ligaments into droplets is delayed, resulting in the formation of a complex filamentous network below the sheet, see insets labelled `zoom' in figure \ref{fig:spray_images}(b). This is also reported by various studies of a wide range of jet and sheet fragmentation processes \citep{negri2010atomization,negri2013spray,negri2017atomization,jung2011experimental,park2008effects,christanti2006quantifying,thompson2007atomization,miller2005collision} and is attributed to the strain hardening properties of polymer solutions \citep{mckinley2005visco}.

\subsection{Drop size}
\label{sec:Drop size}

\begin{figure}
  \centerline{\includegraphics[scale=1]{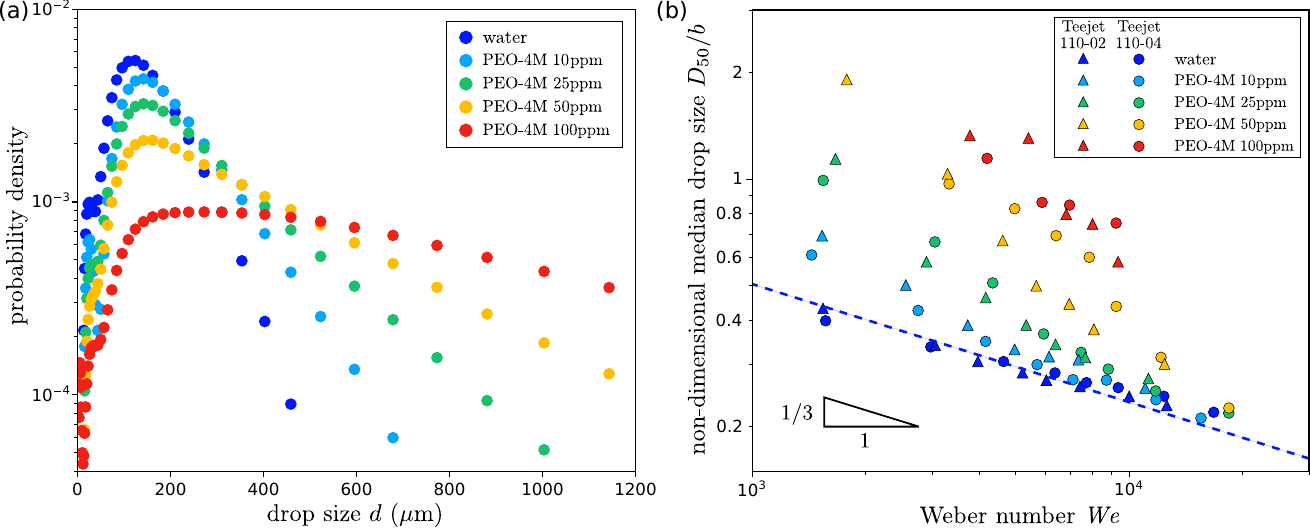}}
  \caption{(a) Drop size distribution measured for sprays of water and PEO-4M solutions of various concentrations obtained from the Teejet 110-02 nozzle at an operating pressure of 3 bars. The distribution is expressed in terms of the probability density function. (b) Non-dimensional median drop size  $D_{50}/b$ against the Weber number $\Web = \rho U_0^2 b /\sigma$ (where $U_0 = Q/A$ is the nozzle velocity) for water and PEO-4M solutions of different concentrations sprayed at different operating pressures from Teejet 110-02 ($\blacktriangle$) and Teejet 110-04 ($\bullet$) nozzles. The dashed line corresponds to equation (\ref{eq:Kooij}) with a prefactor $c_1 = 1.65$. Due to increasing polymer degradation with increasing flow rate, the relaxation time $\tau$ decreases with increasing Weber number for a given polymer concentration, see \S\ref{sec:Mechanical degradation}.}
\label{fig:drop_size}
\end{figure}

We now turn to the size of spray droplets as measured by laser diffraction. In figure \ref{fig:drop_size}(a), we compare the drop size distribution measured for water and PEO-4M solutions of various concentrations for a given nozzle and operating pressure. The flow rate is the same to within 6\% for all these measurements since its dependence on operating pressure is only marginally influenced by polymer addition for flat fan nozzles. The figure shows that spray droplets of polymer solutions are overall bigger than for the water solvent alone and feature a broader size distribution. We first focus on the characteristic drop size, leaving discussions of the distribution to \S\ref{sec:Drop size distribution}.

We choose the volume median diameter $D_{50}$ as a characteristic drop size. In figure \ref{fig:drop_size}(b), we plot $D_{50}$ rescaled by the thickness $b$ of the nozzle opening against the Weber number 
\begin{equation}
\Web = \frac{\rho U_0^2 b}{\sigma}
\label{eq:We}
\end{equation}
for sprays of water and PEO-4M solutions of various concentrations and for all nozzles. The data in figure \ref{fig:drop_size}(b) is obtained by varying the liquid flow rate for each polymer concentration and nozzle. In the case of water sprays, the decrease with increasing Weber number is well captured by equation (\ref{eq:Kooij}), corresponding to the scaling of \citet{kooij2018determines} for Newtonian liquids. Polymer addition to the water solvent results in larger values of $D_{50}$. Figure \ref{fig:drop_size}(b) suggests that the effect of viscoelasticity is stronger at lower liquid velocities $U_0$, while the median drop size of polymer sprays seems to converge asymptotically towards the Newtonian behaviour at high velocity. However, this can be partially due to mechanical degradation since, for a given polymer concentration and nozzle, the extensional relaxation time $\tau$ decreases as $U_0$ increases, see \S\ref{sec:Mechanical degradation}.

\subsection{Universal scaling of the median drop size}
\label{sec:Universal scaling of the median drop size}

\begin{figure}
  \centerline{\includegraphics[scale=1]{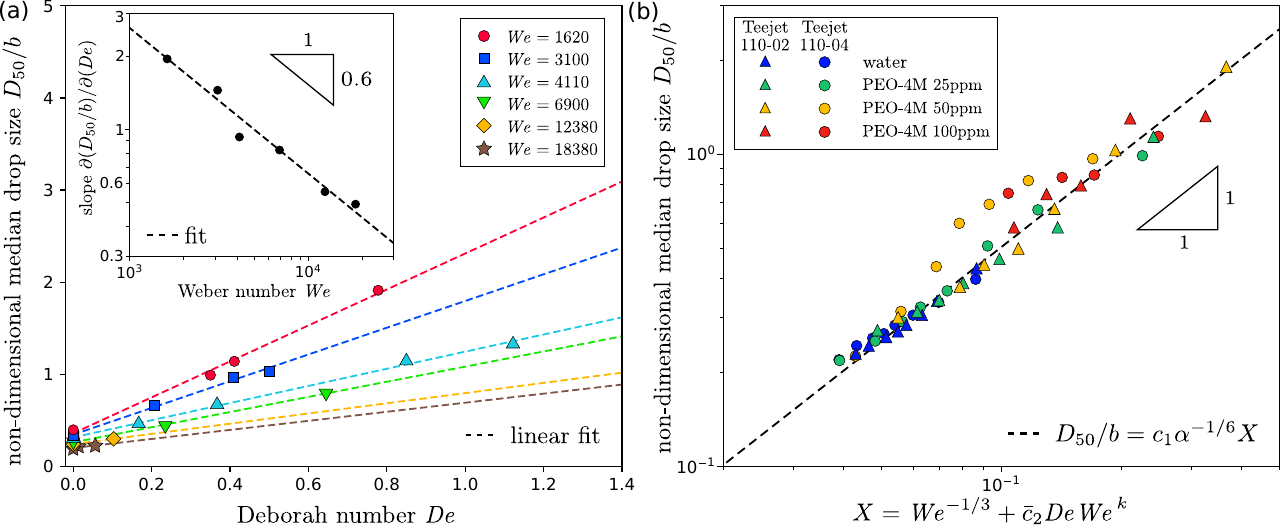}}
  \caption{(a) Non-dimensional median droplet size $D_{50}/b$ against the Deborah number $\Deb = \tau/(\rho b^3 / \sigma)^{1/2}$ for six sets of data points. Within each set, the Weber number $\Web = \rho U_0^2 b /\sigma$ is the same to within up to $15$\%. Dashed lines correspond to the best linear fits. Inset: slope of the linear fits against $\Web$. The dashed line correspond to the best power law fit with exponent $k = -0.6$. (b) $D_{50}/b$ against $\Web^{-1/3} + \bar{c}_2 \Deb \Web^{k}$, with $k = -0.6$ and $\bar{c}_2 = 33$, for the data of figure \ref{fig:drop_size}(b). The dashed line corresponds to equation (\ref{eq:master_curve}) with $c_1 = 1.65$.}
\label{fig:master_curve}
\end{figure}

The zero-shear viscosity of polymer solutions used in spray experiments reaches $1.28$~mPa\,s at 100 ppm. Hence, the increase in shear viscosity cannot account for the significant increase in median drop size observed in figure \ref{fig:drop_size}(b) since we know from \citet{kooij2018determines} that sprays of Newtonian liquids of viscosity up to $32.3$~mPa\,s exhibit nearly identical median drop sizes as those of water sprays. The increase in $D_{50}$ can hence be attributed to elastic effects, which we choose to quantify using the extensional relaxation time $\tau$ measured from filament thinning (CaBER) measurements. Viscoelastic effects are expected to be significant when $\tau$ is comparable to the inertio-capillary Rayleigh time scale $\sqrt{\rho b^3/\sigma}$, i.e. when the Deborah number defined as
\begin{equation}
\Deb = \frac{\tau}{\sqrt{\rho b^3 / \sigma}}
\label{eq:De}
\end{equation}
is of order unity. Note that this definition is different from the local Deborah number considered by \citet{keshavarz2015studying} in air-assisted jet atomization, where the size of the ligaments was chosen as the characteristic length scale.

In order to bypass the polymer degradation problem discussed in \S\ref{sec:Mechanical degradation} and obtain the true physical dependence of the median drop size on the Weber and Deborah numbers, we select data points from figure \ref{fig:drop_size}(b) corresponding to the same Weber number and plot the associated non-dimensional median drop size $D_{50}/b$ against the Deborah number $\Deb$ (figure \ref{fig:master_curve}(a)). We account for mechanical degradation by choosing the relaxation time of the degraded solution samples collected from the spray. We observe a linear dependence of $D_{50}/b$ on $\Deb$ where the value at $\Deb = 0$ corresponds to the Newtonian value. The slope of this linear dependence decreases with increasing Weber number. The inset in figure \ref{fig:master_curve}(a) shows that it follows a power law $\Web^{k}$ with an exponent $k = -0.6 \pm 0.04$. We can therefore write the following empirical formula for the the median drop size: 
\begin{equation}
D_{50}/b = c_1 \alpha^{-1/6} \left( \Web^{-1/3} + \bar{c}_2 \Deb \Web^{k} \right),
\label{eq:master_curve}
\end{equation}
where we find $\bar{c}_2 = 33$ for $k = -0.6$. Note that $\bar{c}_2$ is not necessarily a purely numerical prefactor as it might depend on the air-liquid density ratio $\alpha$. This expression reduces to equation (\ref{eq:Kooij}) for Newtonian liquids ($\Deb = 0$). We show in figure \ref{fig:master_curve}(b) that this empirical expression leads to a collapse of the data of figure \ref{fig:drop_size}(b) on a single master curve. Experiments involving other types of viscoelastic liquids would be necessary to investigate the universality of this scaling; we limit ourselves here to rather dilute solutions of high-molecular-weight flexible polymers. 

\subsection{General expression of the median drop size}
\label{sec:General expression of the median drop size}

\begin{figure}
  \centerline{\includegraphics[scale=1]{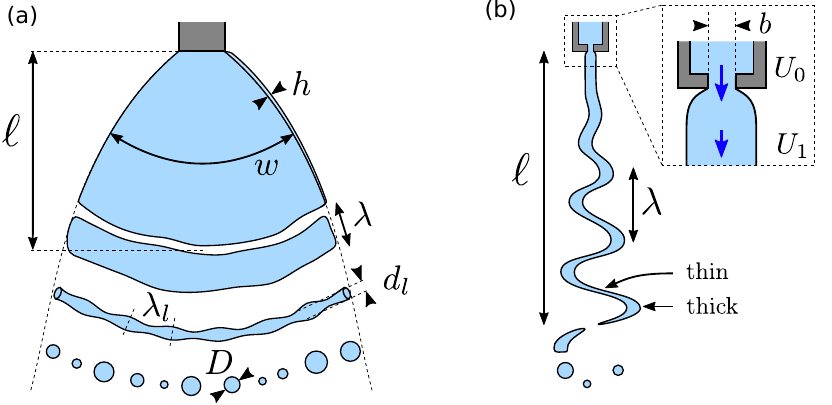}}
  \caption{Schematics of sheet disintegration in (a) front and (b) side view.}
\label{fig:drawing}
\end{figure}

We now address the question of the physical origin of the increasing size of spray droplets with increasing liquid elasticity. To this end, we first derive a general expression of the characteristic drop diameter $D_{50}$ in terms of quantities that do not depend on the specific liquid rheology.

Figure \ref{fig:drawing} provides a schematic representation of the sheet fragmentation mechanism in front view (a) and side view (b). We recall that the sheet fragmentation is initiated by the growth of Squire waves along the sheet. This flapping motion causes the sheet to break up at a radial distance $l$ from the nozzle, henceforth referred to as the breakup length, leaving sheet fragments which then retract into ligaments due to surface tension. Since the size of these sheet fragments is of the order of the wavelength $\lambda$ of the Squire instability, we can use mass conservation to find that the characteristic diameter $d_l$ of the ligaments should scale as
\begin{equation}
  d_l \sim \sqrt{\lambda h(l)},
  \label{eq:d_ligament}
\end{equation}
where $h(r)$ is the sheet thickness at a radial distance $r$ from the nozzle. If the liquid velocity is constant along the sheet, flow rate conservation gives
\begin{equation}
  h = \xi A / w,
  \label{eq:h_w}
\end{equation}
where $\xi = U_0/U_1$ is the ratio between the liquid velocity $U_0 = Q/A$ in the nozzle of area $A$ and the liquid velocity $U_1$ in the sheet, and where $w$ is the length of the arc of a circle of radius $r$ stopping at the sheet edges (see figure \ref{fig:drawing}). We expect $w = \theta r$ for a sheet expanding with a constant angle $\theta$. This relation neglects the presence of rims at the edges of the sheet, which are observable in the front images of figure \ref{fig:spray_images}. Finally, the characteristic drop diameter $D_{50}$ can be deduced from the characteristic ligament diameter using mass conservation, which gives 
\begin{equation}
  D_{50} \sim d_l \, (\lambda_l/d_l)^{1/3},
  \label{eq:D_d}
\end{equation}
where $\lambda_l$ is the characteristic distance between successive drops along the ligament. If drops are formed regularly along the ligament, $\lambda_l$ is the wavelength of the Rayleigh-Plateau instability, which is $\lambda_l/d \approx 4.55$ for Newtonian liquids in the absence of viscous effects.

Equations (\ref{eq:d_ligament})--(\ref{eq:D_d}) do not depend on the rheology of the liquid being sprayed. Therefore, we can measure how different quantities such as $\lambda$, $l$, $\xi$ and $w$ are affected by polymer addition and hence identify which effects play an important role in explaining the increase in drop size with increasing polymer concentration. Measurements of these quantities are reported in \S\S\ref{sec:Die swell}--\ref{sec:Squire wavelength}.

\subsection{Die swell}
\label{sec:Die swell}

\begin{figure}
  \centerline{\includegraphics[scale=1]{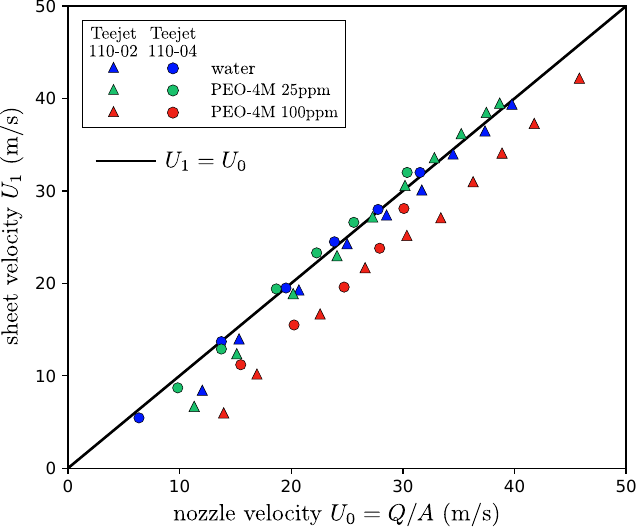}}
  \caption{(a) Velocity $U_1$ of the liquid in the unbroken part of the sheet as a function of velocity $U_0$ of the liquid in the nozzle. The line indicates $U_1=U_0$. The sprays are made of water and PEO-4M solutions of different concentrations and are generated at different operating pressures from both nozzles.}
\label{fig:U1}
\end{figure}

Viscoelastic liquids extruded from orifices are known to potentially expand in air due to the well known die swell effect \citep{tanner1970theory,tanner2005theory}. In our experiments, this would cause spray sheets of polymer solutions to be thicker than water sheets, potentially leading to the formation of larger spray droplets. If die swell occurs, we expect to observe a reduction of the liquid velocity $U_1$ in the sheet compared to the liquid velocity $U_0=Q/A$ in the nozzle, as shown schematically in figure \ref{fig:drawing}(b). Therefore, in order to verify if die swell occurs, we measure the radial liquid velocity $U_1$ in the spray sheet by a particle tracking method, using silver-coated hollow glass spheres of diameter 100 $\mu$m as tracers.

We find that $U_1$ is constant along the unbroken part of the sheet, as expected if gravitational acceleration is negligible, which is the case since $U_1/\sqrt{gl} > 10$ in our experiments. $U_1$ is plotted in figure \ref{fig:U1} as a function of $U_0$ for sprays of water and of $25$ and $100$~ppm PEO-4M solutions formed using both nozzles. We find that $U_1 = U_0$ for both water and the $25$~ppm solution, which means that there is no swelling. For the $100$~ppm solution however, we observe that $U_1$ is smaller than $U_0$ with a constant shift of about 6~m/s, meaning that the velocity ratio $\xi = U_0/U_1$ decreases as $U_0$ increases. This is opposite to the trend expected from the elastic die-swell theory of \citet{tanner1970theory,tanner2005theory}. This is most likely due to the reduction of the liquid elasticity with increasing flow rate caused by mechanical degradation, as discussed in \S\ref{sec:Mechanical degradation}. 

We conclude from these measurements that die swell cannot be the main cause of the increasing droplet size in sprays since droplets of $25$~ppm PEO-4M solutions are significantly larger than water droplets while no die swell is observed at this concentration, see figure \ref{fig:drop_size}(b). 

\subsection{Breakup length}
\label{sec:Breakup length}

\begin{figure}
  \centerline{\includegraphics[scale=1]{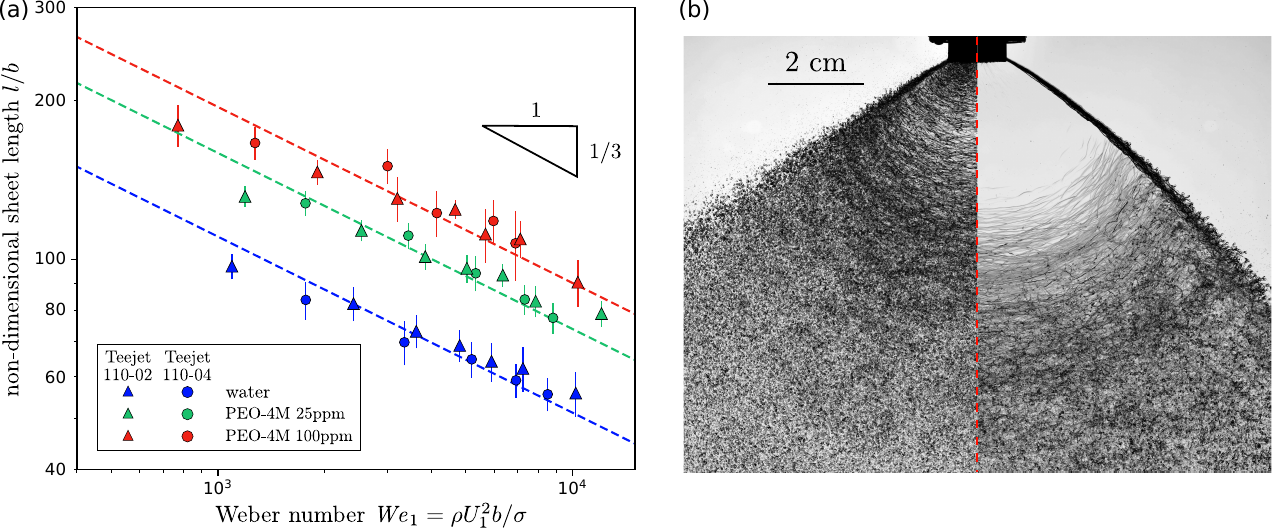}}
  \caption{(a) Non-dimensional sheet length $l/b$ as a function of the Weber number $We_1 = \rho U_1^2 b /\sigma$ calculated from the liquid velocity $U_1$ in the sheet. Lines indicate best fits of equation (\ref{eq:l_Newtonian}) with different prefactors. The sprays are made of water and PEO-4M solutions of different concentrations and are generated at different operating pressures from both nozzles. (b) Juxtaposition of 30 different front-view images of a spray of water (left half) and of 25\,ppm PEO-4M solution (right half). The sprays are generated by the Teejet 110-04 flat fan nozzle at an operating pressure of 2 bar.}
\label{fig:breakup_length}
\end{figure}

The breakup length $l$ also plays a role in setting the median drop size according to \S\ref{sec:General expression of the median drop size}. In our experiments, values of $l$ are estimated from front-view images of the sheet similar to figure \ref{fig:spray_images} obtained from high-speed photography. Average values are obtained from 30 pictures for each set of parameters. Results are presented in figure \ref{fig:breakup_length}(a) where $l/b$ is plotted against the Weber number $We_1 = \rho U_1^2 b /\sigma$ based on the liquid velocity in the sheet, for sprays of water and of $25$ and $100$~ppm PEO-4M solutions formed using both nozzles. We observe that $l$ increases significantly with polymer concentration.

For water, we note that experimental values of $l$ in figure \ref{fig:breakup_length}(a) are consistent with the inviscid Newtonian theory of \citet{kooij2018determines}, which predicts 
\begin{equation}
  l/b \sim \alpha^{-2/3} \Web_1^{-1/3},
  \label{eq:l_Newtonian}
\end{equation}
similar to the scaling of the radius of a circular flapping liquid sheet \citep{villermaux2002life}. Surprisingly, the dependence on the Weber number is also close to a power law with exponent $-1/3$ for the two polymer concentrations in figure \ref{fig:breakup_length}(a), at least at high Weber numbers. However, this may not be the true physical $\Web$ exponent. Indeed, in this case, the decrease of the sheet length with liquid velocity may be partly attributed to the reduction of the liquid elasticity caused by mechanical degradation, see \S\ref{sec:Mechanical degradation}.

According to equations (\ref{eq:d_ligament}-\ref{eq:D_d}), an increase in breakup length alone, while other parameters remain constant, would lead to smaller droplets. A possible compensating effect is the narrowing of the sheet in the orthoradial direction observed for viscoelastic spray sheets. This effect is illustrated in figure \ref{fig:breakup_length}(b) in which we juxtapose 30 images of a spray of water (left half) and of 25~ppm PEO-4M solution (right half), corresponding to the same nozzle and flow rate, to obtain a better visualization of the edges of the spray sheet. This figure reveals that the water sheet expands linearly with a constant angle $\theta$, i.e. $w(r) = \theta r$, while the viscoelastic sheet exhibits a concave profile with $w(r) < \theta r$. This effect is most likely a consequence of the elastic stresses arising from the stretching of polymer molecules caused by the sheet expansion. However, measurements of the $w(r)$ profile reveal that the value of $w(r=l)$ actually increases with polymer concentration because of the significant increase in breakup length.

Using the measured values of $\xi$ and of $w(l)$, we find that the sheet thickness $h(l)$ at breakup, calculated using equation (\ref{eq:h_w}), doesn't increase significantly with increasing polymer concentration. Hence, none of the effects discussed so far can explain the increase in spray droplet size. 

Finally, we turn our attention to the wavelength $\lambda$ of the Squire instability responsible for the sheet fragmentation.

\subsection{Squire wavelength}
\label{sec:Squire wavelength}

\begin{figure}
  \centerline{\includegraphics[scale=1]{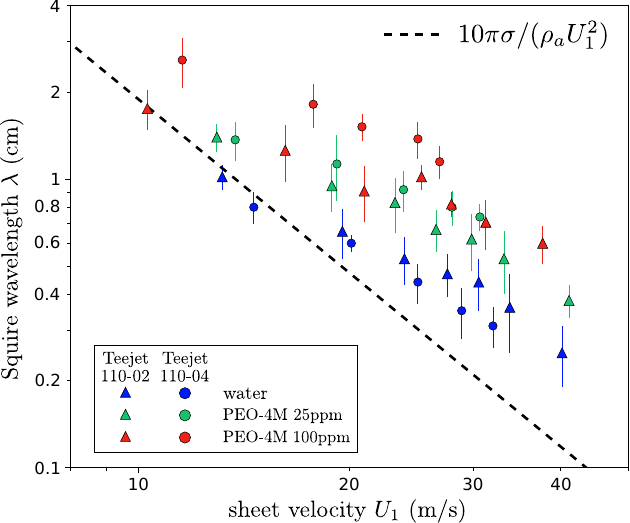}}
  \caption{Squire wavelength $\lambda$ as a function of the liquid velocity $U_1$ in the sheet. $\lambda$ is estimated from high-speed pictures of the spray, see e.g. figure \ref{fig:spray_images}(b). The sprays are made of water and PEO-4M solutions of different concentrations and are generated at different operating pressures from both nozzles. The dashed line corresponds to the prediction of equation (\ref{eq:lambda}).}
\label{fig:lambda}
\end{figure}

The Squire wavelength, which sets the size of sheet fragments detaching from the sheet, also plays a role in setting the median drop size according to \S\ref{sec:General expression of the median drop size}. Values of $\lambda$ are estimated from front-view images of the sheet (cf. figure \ref{fig:spray_images}) obtained from high-speed photography. Average values are obtained from 30 images for each set of parameters. Results are presented in figure \ref{fig:lambda}(b), where $\lambda$ is plotted against the liquid velocity $U_1$ for sprays of water and of $25$ and $100$~ppm PEO-4M solutions formed using both nozzles. These values of $\lambda$ are in agreement with values estimated from side-view images of the spray. We observe that $\lambda$ increases with polymer concentration, which suggests an increase of the size $d_l$ of the ligaments resulting from the retraction of sheet fragments according to equation (\ref{eq:d_ligament}). 

For Newtonian inviscid sheets in still air, the Squire wavelength is \citep{squire1953investigation} 
\begin{equation}
  \lambda = 10 \pi \frac{\sigma}{\rho_{a} U_1^2},
  \label{eq:lambda}
\end{equation}
with $\rho_{a}$ the air density and prefactor $10 \pi$ as derived by \citet{villermaux2002life} for circular sheets. Comparing water data in figure \ref{fig:lambda}(b) with equation (\ref{eq:lambda}), we find that experimental values are underpredicted by a factor up to 2 at high liquid velocities. A possible explanation for this difference is air entrainment along the liquid flow, yielding a velocity difference smaller than $U_1$ across the air-liquid interface.


\subsection{Physical origin of the increase in drop size}
\label{sec:Physical origin of the increase in drop size}

\begin{figure}
  \centerline{\includegraphics[scale=1]{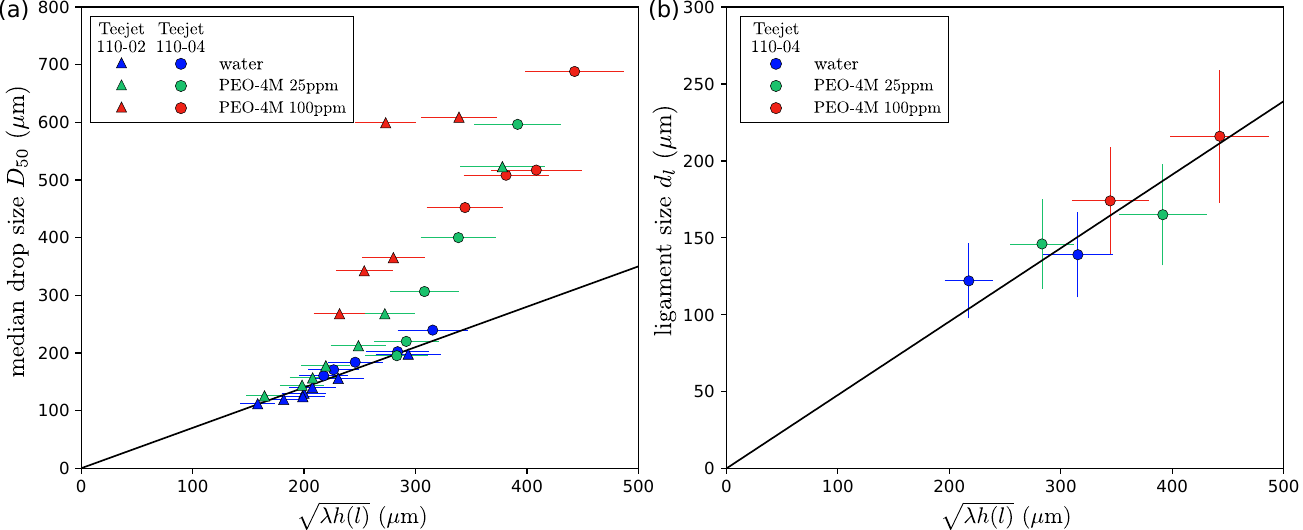}}
  \caption{(a) Median droplet size $D_{50}$, measured from laser diffraction, against $\sqrt{\lambda h(l)}$ where $\lambda$ is the measured Squire wavelength and where the sheet thickness $h(l)$ at breakup is calculated from equation (\ref{eq:h_w}) with measured values of $\xi$ and $w(l)$. The solid line corresponds to the best linear fit for water with slope 0.70. (b) Characteristic diameter $d_l$ of the ligaments, measured from high speed images of the spray, against $\sqrt{\lambda h(l)}$. The solid line corresponds to the best linear fit with slope 0.48. The sprays are made of water and PEO-4M solutions of different concentrations and are generated at different operating pressures from both 110-02 and 110-04 nozzles in (a) and only the 110-04 nozzle in (b).}
\label{fig:D_formula}
\end{figure}


We can now check if the increase in Squire wavelength accounts for the increase in droplet size with increasing viscoelasticity. As equation (\ref{eq:d_ligament}) predicts that the ligament diameter scales with $\sqrt{\lambda h(l)}$, in figure \ref{fig:D_formula}(a), we plot the median drop size $D_{50}$ against $\sqrt{\lambda h(l)}$, where $\lambda$ is the measured Squire wavelength and sheet thickness $h(l)$ at breakup is calculated from equation (\ref{eq:h_w}) with measured values of $\xi$ and $w(l)$. We obtain a relatively good collapse of data points for the different liquids and nozzles, which would not have been possible if the increase of $\lambda$ with polymer concentration had been neglected. Hence, the increase in Squire wavelength is a key effect for explaining the increase in droplet size. For comparison, we find that the effect of elasticity on $h(l)$ is less important since assuming $h(l) \sim A/l$ has less impact on the quality of the collapse. 

However, we also find that $D_{50}$ is only proportional to $\sqrt{\lambda h(l)}$ for water in figure \ref{fig:D_formula}(a), while the linear fit obtained for water underestimates the droplet size by a factor of up to 2 for polymer solutions exhibiting the largest relaxation times. To investigate this further, we measure the characteristic diameter $d_l$ of the ligaments formed in the sheet breakup region from high-speed images of the spray. In figure \ref{fig:D_formula}(b) we plot $d_l$ against $\sqrt{\lambda h(l)}$ for water and polymer solutions. The data reveal that $d_l$ is proportional to $\sqrt{\lambda h(l)}$ with a numerical prefactor $0.48$, consistent with equation (\ref{eq:d_ligament}). Therefore, according to figure \ref{fig:D_formula}(a), the ratio $D_{50}/d_l$ increases with increasing viscoelasticity.

We conclude that the increase in droplet size in ligament-mediated fragmentation of a viscoelastic sheet is partially explained by an increase in ligament diameter, caused by an increase of the Squire wavelength, which sets the size of the sheet fragments detaching from the sheet. We hypothesize that the increasing discrepancy between droplet and ligament diameters observed for increasing viscoelasticity is caused by an increasing distance $\lambda_l$ between droplets along a ligament, resulting in the formation of fewer, bigger droplets according to equation (\ref{eq:D_d}). 

\subsection{Droplet size distribution}
\label{sec:Drop size distribution}

\begin{figure}
  \centerline{\includegraphics[scale=1]{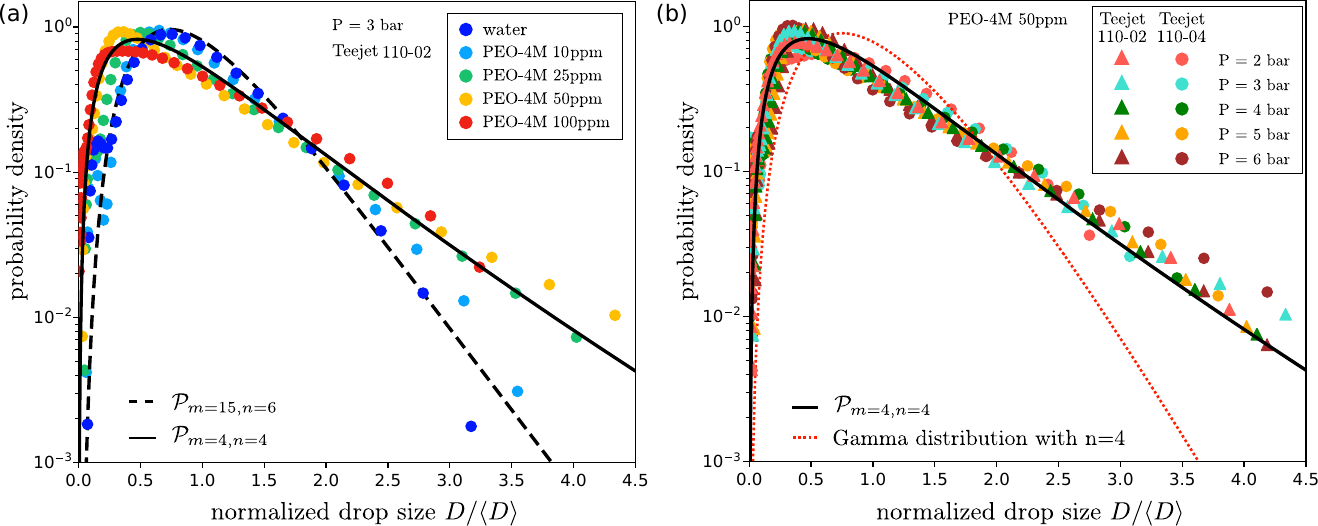}}
  \caption{(a) Droplet size distributions of figure \ref{fig:drop_size}(a) rescaled by the mean droplet diameter $\langle D \rangle$ for sprays of water and PEO-4M solutions of various concentrations obtained from the Teejet 110-02 nozzle at an operating pressure of 3 bar. The dashed and solid lines correspond to the best fits of the compound gamma distribution (equation~(\ref{eq:compound_gamma})) for water ($m=15$, $n=6$) and for high polymer concentrations ($m=4$, $n=4$), respectively. (b) Droplet size distributions for sprays of a 50~ppm PEO-4M solution generated at different operating pressures from both nozzles, covering a range of Weber numbers $We$ from 3000 to 9000 and a range of Deborah numbers $De$ from 0.15 to 0.5. The solid line corresponds to the best fit of equation (\ref{eq:compound_gamma}) ($m=4$, $n=4$). The dashed line corresponds to the gamma distribution (equation~(\ref{eq:gamma})) for $n=4$.}
\label{fig:PDF_Antoine}
\end{figure}

After having discussed the increase in median droplet size with increasing viscoelasticity, we now turn our attention to the distribution of droplet sizes in the spray. Droplet size distributions are presented in figure \ref{fig:PDF_Antoine}(a) in terms of normalized probability density functions (PDFs) for sprays of water and PEO solutions of different concentrations, generated by the Teejet 110-02 nozzle at a fixed operating pressure of $3$~bar. The data are the same as in figure \ref{fig:drop_size} but plotted against the droplet diameter $D$ rescaled by the average droplet diameter $\langle D \rangle$. We observe that adding polymers to the water solvent results in the formation of a broader range of droplet sizes, as illustrated by the shift of the PDF peak to the left and by the increase in PDF values in the `tail' of the distribution. In addition, we observe no significant change for the largest concentrations, suggesting a saturation towards a universal distribution. 

\citet{kooij2018determines} and \citet{villermaux2011drop} showed that PDFs obtained for the fragmentation of Newtonian liquid sheets are accurately captured by the compound gamma distribution 
\begin{equation}
  \mathcal{P}_{m,n}\left( x = \frac{D}{\langle D \rangle} \right) = \frac{2 ( mn )^{\frac{m+n}{2}}  x ^{\frac{m+n}{2}-1}}{\Gamma(m)\Gamma(n)} \mathcal{K}_{m-n}\left(2\sqrt{mnx}\right),
  \label{eq:compound_gamma}
\end{equation}
where $\mathcal{K}$ is the modified Bessel function of second kind and $\Gamma$ is the gamma function. The parameters $n$ and $m$ correspond to the corrugation of the ligaments formed during sheet breakup (from which droplets are formed) and the width of the ligament diameter distribution, respectively. In the limit $m \to \infty$, where all ligaments have the same diameter, equation (\ref{eq:compound_gamma}) reduces to the gamma distribution 
\begin{equation}
 \Gamma \left(n, x = \frac{D}{\langle D \rangle} \right) = \frac{n^n}{\Gamma(n)} x^{n-1} e^{-nx}, 
  \label{eq:gamma}
\end{equation}
which corresponds to a fragmentation-coalescence scenario \citep{villermaux2007fragmentation}. Large values of $n$ correspond to smooth ligaments producing droplets of the same size.

We find that the PDFs in figure \ref{fig:PDF_Antoine}(a) are best fitted by the compound gamma distribution of equation (\ref{eq:compound_gamma}) with values $m=15$ and $n=6$ for water and with lower values $m=4$ and $n=4$ for the highest PEO concentrations. This means that the presence of polymers results in the formation of more corrugated ligaments with a broader ligament diameter distribution. 

The same low values of $m=4$ and $n=4$ are also found to yield the best fit of the PDFs in figure \ref{fig:PDF_Antoine}(b), corresponding to different operating pressures for a fixed polymer concentration (50~ppm) for both flat fan nozzles. The data in this figure cover Weber numbers $We$ ranging from 3000 to 9000 and Deborah numbers $De$ ranging from 0.15 to 0.5, which reinforces the idea of a saturation towards a universal distribution when the liquid is sufficiently viscoelastic, independent of Weber and Deborah numbers.

These results are in part consistent with the work of \citet{keshavarz2016ligament} who showed that the corrugation of viscoelastic ligaments, produced from various fragmentation processes, saturates with increasing viscoelasticity to a low value $n = 4$. The authors show that this value corresponds to a geometrical limit which can not be exceeded in ligament-mediated fragmentation phenomena. The authors report experimental PDFs which are well fitted by the gamma distribution (\ref{eq:gamma}) with $n=4$ for three different fragmentation processes: air-assisted jet atomization, oblique jet impact atomization and droplet impact on a small target (no flat fan nozzles). This means that ligaments were monodisperse in their experiments ($m \to \infty$). However, this is not the case in our measurements using flat fan nozzles, as illustrated in figure \ref{fig:PDF_Antoine}(b) where a gamma distribution with $n=4$ is shown to significantly underpredict the width of droplet size distribution. This discrepancy shows that the droplet size distribution obtained from the fragmentation of viscoelastic liquids depends on the details of the fragmentation process: different spraying techniques may lead to different widths of ligament size distributions, i.e. different values of $m$.

\section{Conclusions}
\label{sec:Conclusions}

The impact of viscoelasticity on sprays produced by agricultural flat fan nozzles was investigated experimentally using dilute aqueous solutions of high-molecular-weight polyethylene oxide. We find that, similarly to Newtonian fluids, the fragmentation of the spray sheet into droplets is initiated by the growth of Squire waves along the sheet, leading to the formation of sheet fragments retracting into ligaments which ultimately break up into droplets. Polymer addition to water was shown to delay the breakup of the sheet to a larger distance from the nozzle and to stabilize the rims of the sheet. In addition, the capillary breakup of ligaments into droplets was delayed by the stabilizing strain-hardening properties of polymer molecules, leading to the formation of a complex filamentous
network as the sheet breaks up. The size distribution of the droplets produced by the spray was measured by laser diffraction at a distance from the nozzle where ligaments had fully broken up into independent droplets. 

As the polymer concentration was increased, we observed the formation of droplets with a higher median size and a broader size distribution. The median droplet size $D_{50}$ was found to increase linearly with the extensional relaxation time $\tau$ of the polymer solution. This relaxation time decreased with increasing liquid velocity for a given polymer concentration due to increasing mechanical degradation. We showed that the median drop size measured for different polymer solutions, sprayed at different operating pressures from different flat fan nozzles, collapses on a single master curve using an empirical expression that relates $D_{50}$ to the Weber and Deborah numbers, see equation \ref{eq:master_curve}). This expression reduces to the Newtonian formula derived by \citet{kooij2018determines} in the absence of viscoelastic effects.

Measurements of different quantities involved in the sheet fragmentation mechanism revealed that the increase in droplet size with increasing polymer concentration is only partly due to the increase of the wavelength of the Squire waves responsible for the sheet breakup; the thickness of the sheet calculated at the point of breakup was not significantly altered by polymer addition. The larger wavelength leads to the formation of larger ligaments from the sheet breakup. However, our results show that this mechanism does not fully explain the measured increase in droplet size, suggesting the existence of a second effect of viscoelasticity in the breakup of the ligaments themselves.

When rescaled by the average drop size, the measured droplet size distributions are well described by a compound gamma distribution where the parameters $n$ and $m$, that give the ligament corrugation and the width of the ligament size distribution, respectively, saturate to values $n=4$ and $m=4$ for the highest polymer concentrations. We can therefore now relate the changes in drop size and its distribution to the presence of polymers, and notably the polymer relaxation time. This then gives a handle for controlling sprays using polymer additives; one important application of this is spray drift reduction in e.g. shipyard or agricultural spraying, where highly toxic anti-fouling or pesticide molecules in small drops are prone to drift away with the wind.

\bibliographystyle{jfm}
\bibliography{bibliography}

\end{document}